\documentclass{article}
\usepackage[margin=1in]{geometry}
\usepackage[utf8]{inputenc}
\usepackage{graphicx}
\usepackage{hhline}
\usepackage{enumitem}
\usepackage{amsmath}
\usepackage{hyperref}
\usepackage{caption}
\usepackage{subcaption}
\usepackage[load=physical,load=synchem]{siunitx}
\usepackage{float}
\usepackage{xcolor}
\usepackage{amsfonts}
\usepackage[normalem]{ulem}
\usepackage{authblk}

\hypersetup{
    colorlinks=true,
    linkcolor=blue,
    filecolor=magenta,      
    urlcolor=cyan,
}

\begin{document}
\title{Design for a \SI{10}{\kilo\electronvolt} Multi-Pass Transmission Electron Microscope}
\author[1]{Stewart A. Koppell}
\author[2]{Marian Mankos}
\author[1]{Adam J. Bowman}
\author[1]{Yonatan Israel}
\author[3,4]{Thomas Juffmann}
\author[1]{Brannon B. Klopfer}
\author[1]{Mark A. Kasevich}
\affil[1]{{\small Physics Department,Stanford University, 382 Via Pueblo Mall, Stanford, California 94305, USA, skoppell@stanford.edu}}
\affil[2]{{\small Electron Optica, 1000 Elwell Court \#110
Palo Alto, California 94303, USA}}
\affil[3]{{\small Faculty of Physics, University of Vienna, A-1090 Vienna, Austria}}
\affil[4]{{\small Department of Structural and Computational Biology, Max F. Perutz Laboratories, University of Vienna, A-1030 Vienna, Austria}}
\date{}
\maketitle

\begin{abstract}
Multi-pass transmission electron microscopy (MPTEM) has been proposed as a way to reduce damage to radiation-sensitive materials. For the field of cryo-electron microscopy (cryo-EM), this would significantly reduce the number of projections needed to create a 3D model and would allow the imaging of lower-contrast, more heterogeneous samples. We have designed a \SI{10}{\kilo\electronvolt} proof-of-concept MPTEM. The column features fast-switching gated electron mirrors which cause each electron to interrogate the sample multiple times. A linear approximation for the multi-pass contrast transfer function (CTF) is developed to explain how the resolution depends on the number of passes through the sample. 
\end{abstract}

\section{Introduction} 
     
Cryo-electron microscopy (cryo-EM) can produce near-atomic-resolution reconstructions of biological molecules by reducing (radiolysis) damage to the sample and  computationally combining a large number ($>10^4$) of low-resolution projections \cite{Henderson1990}. While the significant product and promise of this technique has inspired rapid development \cite{Agard2014}, there are important applications which it cannot address.

Even anticipating future advances in hardware (better detectors, reliable phase plates, improved sample preparation) and analysis methods, it seems unlikely conventional Cryo-EM will achieve atomic or near-atomic resolution of individual particles \cite{Glaeser2015}. In order for ensemble approaches to be effective, the molecules must be large enough to produce sufficient contrast so they can be accurately localized (the smallest molecule imaged to date weighs \SI{60}{\kilo\dalton} \cite{Khoshouei2017}). In some important applications, large ensembles of identical specimens may not be available (e.g. protein dynamics in solution, intrinsically disordered proteins, polymers).

In order to move beyond these limitations, it will be necessary to improve the dose-limited resolution (DLR) \cite{Egerton2014} of transmission electron microscopes (TEMs). Methods for improving DLR include adaptive or compressive imaging \cite{Okamoto2010}, entangled electrons to overcome shot noise statistics\cite{Yurke1986b}\cite{okamoto2014}, or interaction-free measurement to obtain silhouettes of the sample \cite{Putnam2009}\cite{Kruit2016}. The microscope described in this paper implements a multi-pass approach (multi-pass TEM or MPTEM) which improves image contrast in bright field, dark field and phase contrast modalities through iterative interactions with the image target. \cite{Juffmann2017}. 

The single-pass shot noise limit can be surpassed by repeatedly interrogating the sample with each probe particle \cite{Luis2002}. This technique can be realized with a sample in a re-imaging optical cavity (where rays retrace their paths each round trip), as has recently been demonstrated in light microscopy \cite{Juffmann2016a}\cite{Klopfer2016}. A multi-pass electron microscope will require a larger excursion from conventional optics. The main challenge is building a re-imaging electron cavity. While linear, electrostatic traps \cite{Zajfman1997} and gated trap end-caps \cite{Fujiwara2004} have been demonstrated, coherent re-illumination of a sample has not. 

In a MPTEM, each electron passes through the sample $m$ times, thereby experiencing an $m$-fold enhanced phase shift and also causing an $m$-fold increase in the expected damage. Assuming the single-pass phase shifts are small, this increases the scattered amplitude by a factor of $m$ and the scattered intensity by a factor of $m^2$. Thus, for dark field imaging, the signal is increased by $m^2$. Assuming that the signal-to-noise ratio (SNR) is limited by electron counting statistics (shot-noise), the noise is increased by $m$. For phase contrast and amplitude contrast (bright field) imaging, the signal is increased by $m$ and the noise is unchanged. For any of these imaging modalities, the damage at constant SNR decreases by a factor of $m$. We can also describe the improvement in terms of the SNR at constant damage, which increases by a factor of $\sqrt{m}$. For large but realistically achievable $m$, this technique can enter an interaction-free regime where there is a significant chance of measuring the phase thickness of a region in the sample without causing any damage at all \cite{Juffmann2017}. A \SI{300}{ \kilo\electronvolt} MPTEM with $m=10$ could carry cryo-EM across the atomic-resolution threshold for ensemble reconstructions and make it possible to image more heterogeneous samples. The increased contrast would significantly reduce the lower limit for particle mass (simulations in \cite{Juffmann2017} show individual alpha helices clearly visible below the critical dose).
     
This paper describes a proof-of-concept design for a \SI{10}{\kilo\electronvolt} microscope. Our design is based on a linear, temporally gated cavity which achieves 4-\SI{6}{\nm} resolution for $m = 10$ passes. In the next sections, we will describe each subsystem of the microscope, starting with an overview of the column and the imaging cycle. An analysis of multi-pass aberrations is presented in section \ref{sec:aberrations}. In section \ref{sec:transient}, we describe simulations of the transient fields in the gated mirror. Finally, in section \ref{sec:performance}, we simulate performance of the microscope imaging a graphene target.
    
\section{Column overview}
    
\begin{figure}
\includegraphics[width=1\linewidth]{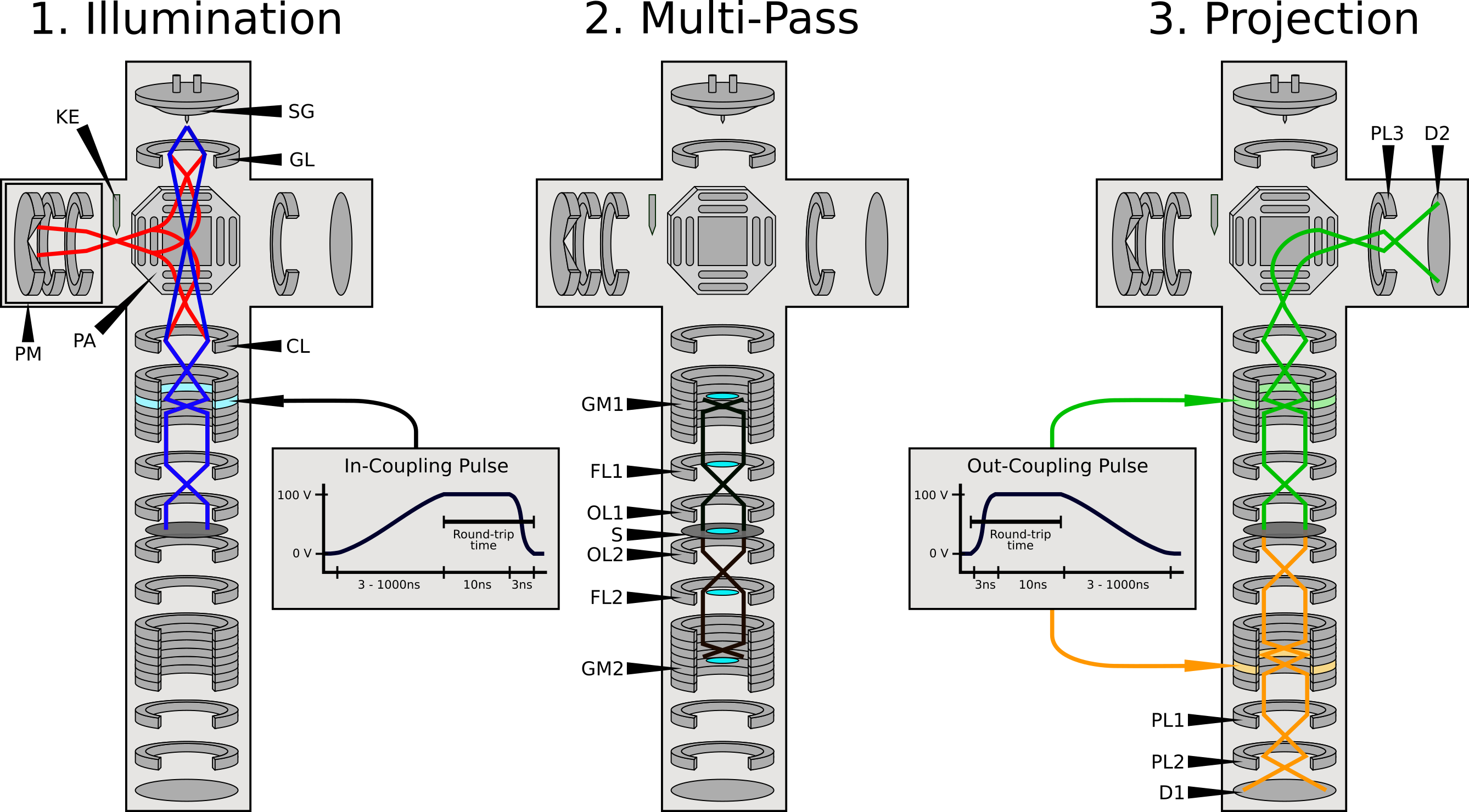}
\caption{Schematic of the MPTEM column, which can be divided into three main systems. The illumination optics (1) couple a wide field coherent beam into the multi-pass optics (2), which cause the beam to repeatedly re-image the sample. Finally, the projection optics (3) receive the beam when it is released from the cavity and magnify it onto one of two detectors. The illumination optics include a Schottky gun (SG), gun lens (GL), and condenser lens (CL). The blue rays show the beam crossovers when the magnetic prism array (PA) is off. When the prism array is on (red rays), the beam is reflected off a mirror (PM) with a knife-edge monochromator (KE). The multi-pass optics include two objective lenses (OL1, OL2), two field lenses (FL1, FL2), and two gated mirrors (GM1, GM2) arranged symmetrically around the sample (S). In order to in- or out-couple electrons the gated mirrors can be quickly switched to strong lenses by applying a 100V RF pulse. If the beam exits through the lower gated mirror, it is magnified by two projection lenses (PL1, PL2) onto detector D1. If the beam exits through the upper gated mirror, it is magnified by the condenser lens, then deflected by the prism array toward a projection lens (PL3) and detector D2.}
\label{fig:schematic}
\end{figure}

Figure \ref{fig:schematic} shows a schematic of the MPTEM. The illumination, multi-passing, and projection optics are used in a three step sequence. First, electrons emitted from a laser-triggered source are focused by the illumination optics into the back focal plane of the upper objective lens (OL1). A magnetic prism array (PA) may be activated to incorporate a mirror monochromator identical to the design described in \cite{Mankos2016}. The illumination must then pass through the upper gated mirror (GM1). In their static configuration, the gated mirrors are ``closed" - they reflect incident electrons. During in-coupling, a low-voltage (\SI{100}{\volt}) nanosecond rise time pulse is applied to an electrode in GM1, temporarily converting it into a strong lens. In order to prevent space-charge effects, the beam current is sufficiently low to make it unlikely that more than one electron will be in-coupled per cycle. During the second step in the imaging sequence, both gated mirrors are closed and the sample is re-imaged $m$ times. In the final step, the beam is out-coupled to the projection optics by ``opening" either GM1 or the lower gated mirror (GM2) with another fast voltage pulse. The sequence is repeated at 1 MHz, so a 1 second measurement could apply 10 electrons per pixel to a 300x300 pixel field of view. 
    
\section{Laser-triggered electron gun}
    
Electron emission must be synchronized with the voltage pulses on the gated mirrors. Electrons which arrive early at the gated mirrors will loose energy to the gating pulse, while those that arrive late will gain energy.
The gun timing requirements depend on the profile of the voltage pulse used to open the gated mirrors. With a Gaussian gating pulse, the gun timing jitter must be less than \SI{100}{\pico\second} to prevent significant broadening of the energy spread of the beam beyond the \SI{0.5}{\electronvolt} thermal spread of the source.  

A laser-triggered Schottky electron gun will be used to generate short ($\sim$\SI{1}{\nano\second}) electron distributions. Femtosecond timing resolution \cite{Hommelhoff2006}\cite{Juffmann2015} and excellent coherence properties \cite{Hommelhoff2015} have been demonstrated for laser-triggered sources and they have been used successfully in TEMs \cite{feist2017}. As an alternative timing strategy, 

Using a flat-top gating pulse profile with Gaussian tails will relax the timing requirement enough to initially test the microscope with a continuous beam current (see section \ref{sec:transient}).

\section{Illumination optics}
 
The illumination optics focus the beam through the (open) GM1 and into the back focal plane of the upper objective lens. With the prism array off, the location of the cross-over between the GL and CL controls the magnification of the illumination system, which determines the size of the illuminated area on the sample. The illumination magnification can be varied by a factor of 50. This allows the field of view to be adjusted between 26 $\mu$m and 0.5 $\mu$m without losing beam current to an aperture. At these upper and lower bounds, the spot size in BFP1 is limited by spherical aberration and the magnified size of the source, respectively. The ideal field of view for high resolution imaging is around 2 microns, where the convergence angle of the illumination will be less than \SI{0.1}{\milli\radian} (full width at half maximum). When using the monochromator, the gun lens establishes a crossover at the knife edge and the field of view is fixed at $\sim$\SI{1.5}{\micron}. 

\section{Multi-pass optics}
    
\begin{figure}
\includegraphics[width=1 \linewidth]{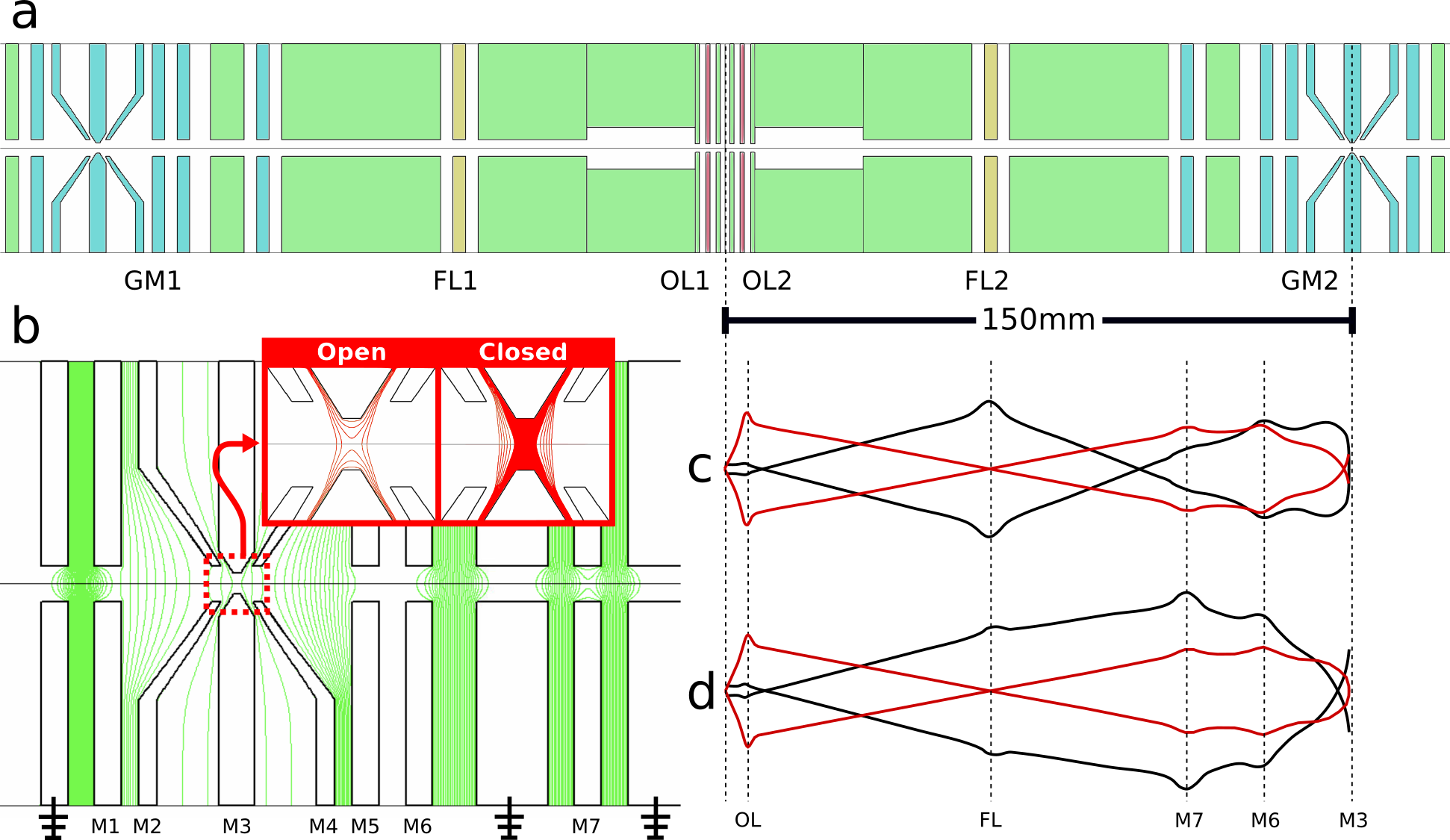}
\caption{Multi-pass optics geometry. a) Full electrode structure with gated mirrors (blue), field lenses (yellow), and objective lenses (red). b) Enlarged view of a gated mirror with equipotential lines spaced by 200V. Electrodes M1-M7 are labeled below. The inset shows equipotential lines spaced \SI{20}{\volt} from \SI{0}{\volt} to \SI{100}{\volt} for open and closed conditions. The solid red region is energetically forbidden. c,d) Paraxial image and field rays traced through half of the multi-pass optics with the same horizontal scale as a). In c) there is a diffraction plane at the turning point, so the image returns to the sample plane inverted. With an image plane at the turning point in d), the sample is re-imaged with the correct orientation but spherical aberration cannot be corrected. }
\label{fig:basic}
\end{figure}

The multi-pass optics consist of two objective lenses (OL1, OL2), two field lenses (FL1, FL2), and two gated mirrors arranged symmetrically around the sample (see figure \ref{fig:basic}). The field lenses are placed at the conjugate image planes established by the objective lenses. 

Two focusing conditions must be enforced by these optics. Rays leaving the sample must return to the same point on the sample (the sample must be re-imaged) and they must return at the correct angle (after passing through an even number of cross-overs, the angles will be reflected across the optical axis). The first focusing condition is enforced by adjusting the focal length of the gated mirrors. The field lenses control the second condition by focusing the field rays into the backfocal plane of each objective lens. Since they are placed at image planes, they only weakly affect image rays. This gives them nearly independent control of the second focusing condition. In practice, the coupling of the field lens potentials to the locations of the image planes is strong enough that precise focusing is an iterative process which involves alternately adjusting the mirror focus and field lens.

The focusing can also be accomplished using a two-lens system between each mirror and objective lens (FL and M7 in figure \ref{fig:basic}). If the mirror focus is always set to infinity (so a collimated input beam remains collimated after reflection), then one degree of freedom in the two-lens system can be permanently fixed. This focusing scheme doesn't require iteration, but is less straight-forward to align.

When both focusing conditions are satisfied, the image (and diffraction) planes for electrons traveling in each direction in the multi-pass optics will coincide. This guarantees that there will either be an image plane or a diffraction plane at the two turning points.

With image planes at the turning points, the beam passes through an even number of cross-overs between interactions with the sample, returning each time with the correct orientation. In this configuration the mirrors can be tuned to correct chromatic aberration but not spherical aberration. With a \SI{3}{\milli\radian} numerical aperture, the image returned to the sample is blurred by \SI{3}{\nm} (the blur from diffraction alone is \SI{2.2}{\nm}, see section  \ref{sec:aberrations}). 

With diffraction planes at the turning points, the mirrors can correct both the spherical and chromatic aberrations of the objective lens, significantly decreasing the round-trip aberrations. In this configuration, the beam passes through an odd number of cross-overs between interactions with the sample, so the image returns inverted. There are a few possible solutions to this problem.

In this case, the sample needs to be confined to one side of the field of view so that the reverse propagating (inverted) image does not interact with the sample. A magnetic dipole field could also be used to cause the inverted image to circumvent the sample. Alternatively, magnetic lenses could be dded to future columns to create a rotation of 180 degrees in each half of the cavity.

The gated mirrors contain 7 electrodes, M1-M7, with 6 electrostatic degrees of freedom (see figure \ref{fig:basic}). M1 is used to position the first image plane outside the multi-pass optics. M2 and M4 share a single potential, $V_{\text{shield}}$. They pinch inward around the pulsed electrode, M3, to minimize the reach of the transient fields (see section \ref{sec:transient}). The main degrees of freedom used to correct aberrations are M5 and the difference between M3 and $V_{\text{shield}}$. M6 is used to set the focal length of the mirror. M7 is an auxiliary lens for testing alternative focusing and aberration-correcting schemes.

\section{Projection optics}
    
During the last round-trip in the multi-pass optics, either GM1 or GM2 is pulsed open. The upper branch of the projection optics is limited to lower resolution imaging and will mainly be used for alignment. If the electron is out-coupled upward, the combined effect of GM1 and the condenser lens causes 90$\times$ magnification. The prism then deflects the electron through a projection lens which adds 2-3$\times$ magnification. 

If the electron is out-coupled downward, it picks up 9x magnification while passing through GM2. The lower projection system applies 10-1000x magnification in two stages and focuses the image onto a YAG scintillator. Since each imaging cycle uses a single electron, detection efficiency is extremely important. At high magnification a \SI{1.3}{\nano\metre} region in the sample is projected onto a 12 micron region at the screen, and a microchannel plate can be used to increase detection sensitivity. When the field of view is \SI{1}{\micro\metre}, the blur from out-coupling (mostly spherical aberrations from OL2 and open GM2) is \SI{2.5}{\nano\metre} at the corner of the image. For a \SI{10}{\micro\metre} field of view, the out-coupling blur increases to \SI{6}{\nano\metre}.

\section{Resolution}\label{sec:aberrations}

Electron mirrors are typically used as aberration correctors, in particular in low energy electron microscopes (LEEMs) \cite{Bauer1994}. Mirrors with four electrodes can correct primary spherical and chromatic aberration \cite{Shao1990} \cite{Preikszas1997}. With five electrodes, they can also correct fifth order spherical aberration \cite{Mankos2012}. In the particular case of a hyperbolic mirror, the aberration coefficients can be calculated analytically \cite{Rempfer1990}, but in general they must be found computationally by scanning large areas of parameter space. 

While the primary purpose of the gated mirrors is to re-image the sample, they can also be used to correct the chromatic and spherical aberrations of the objective lens each pass. Round-trip aberrations were minimized using Munro's Electron Beam Software (MEBS). Chromatic aberration can be fully corrected, leaving the spherical aberration of the objective lenses ($C_s\sim$ \SI{80}{\milli\metre}) as the dominant source of image blur. Fifth order geometric aberrations and fourth and fifth rank chromatic aberrations are negligible. Calculating the resolution of the full multi-pass imaging process is difficult due to the non-linearity of the multi-pass contrast transfer function (CTF). Rather than develop a full treatment of the multi-pass CTF in this paper, we will motivate a linear approximation from which we will infer how resolution scales with $m$. 

In the analysis which follows, we assume the single pass scattering probability $\alpha^2$ is small enough that a linear approximation can be used. For plane wave illumination, we can write the 1-pass exit wavefunction as
\begin{equation}
    \psi_{1}(\vec{q})=(1-\alpha)\delta(\vec{q}) + \alpha P(\psi_s(\vec{q}))
\end{equation}
 where $\vec{q}$ is the spatial frequency, $\delta(\vec{q})$ un-scattered (plane) wave, $\psi_s(\vec{q})$ is the normalized scattered wave, and $P$ is the pupil function
 \begin{equation}
     P(\psi(\vec{q}))=A(\vec{q})e^{i\chi(\vec{q})}\psi(\vec{q})
 \end{equation}
where $A(\vec{q})$ is the aperture function and $\chi(\vec{q})=\pi\lambda\Delta f|\vec{q}|^2+\frac{1}{2}\pi\lambda^3C_s|\vec{q}|^4$ is the phase shift from spherical aberration (with coefficient $C_s$) and defocus ($\Delta f$) with electron wavelength $\lambda$. After $m$ passes, the exit wavefunction is

\begin{equation}
    \psi_{m}(\vec{q})=(1-\alpha)^{m}\delta(\vec{q}) + \alpha(1-\alpha)^{m-1}\sum_{n=1}^mP^n(\psi_s(\vec{q})) +\mathcal{O}(\alpha^2)
\end{equation}

The second term can be interpreted as the sum of the waves which scatter once (at pass $m-n+1$) and then interact only with the optical aberrations and the aperture for the remaining passes. The excluded higher order terms can be written as the sum of convolutions of $P^n(\psi_s)$. The intensity at the detector is

\begin{align}
    I_m(\vec{q})&=(1-2m\alpha)|\delta(\vec{q})|^2+2 \alpha\text{Re}\left(e^{i\phi(\vec{q})}\delta(\vec{q})*\sum_{n=1}^mP^n(\psi_s(\vec{q}))\right)+\mathcal{O}(\alpha^2)
\end{align}
where $*$ denotes convolution and $\phi(\vec{q})$ is a phase shift added in the projection optics to produce phase contrast. When using a phase plate,  $\phi(\vec{q})=\frac{\pi}{2}\delta(\vec{q})$. For defocus contrast, $\phi(\vec{q})=\pi\lambda\Delta f_p|\vec{q}|^2$. The terms in this series are each of the once-scattered waves interfering with the un-scattered wave. The interference among once-scattered waves is $\mathcal{O}(\alpha^2)$, the same order as the first non-linear terms (twice-scattered waves interfering with the un-scattered wave). We will use a hard aperture: $A(\vec{q})=0$ for $q<q_\text{max}$ and $A(\vec{q})=1$ otherwise. Then weak-scattering multi-pass CTF, normalized to the one-pass CTF, is

\begin{align}
    \text{MPCTF}_m(\vec{q})&=\text{Im}\left(A(\vec{q})e^{i\phi(\vec{q})}\sum_{n=1}^me^{in\chi(\vec{q})}\label{eq:MPCTF}\right)\\&=A(\vec{q})\sum_{n=1}^m\sin\left(n\chi(\vec{q})+\phi(\vec{q})\right)
\end{align}

In this linear approximation, the MPCTF is the sum of 1-pass CTFs from the waves scattered from each pass. A more complete study of the MPCTF would include the effect of spatial and temporal coherence envelopes, which are important for calculating contrast at high spatial frequencies. However, multi-pass contrast is significantly diminished for spatial frequencies higher than the point resolution, as the scattered waves from each pass no longer add constructively. We will leave multi-pass envelope functions for a future analysis and focus on understanding the point resolution.

\begin{figure}
\includegraphics[width=1\linewidth]{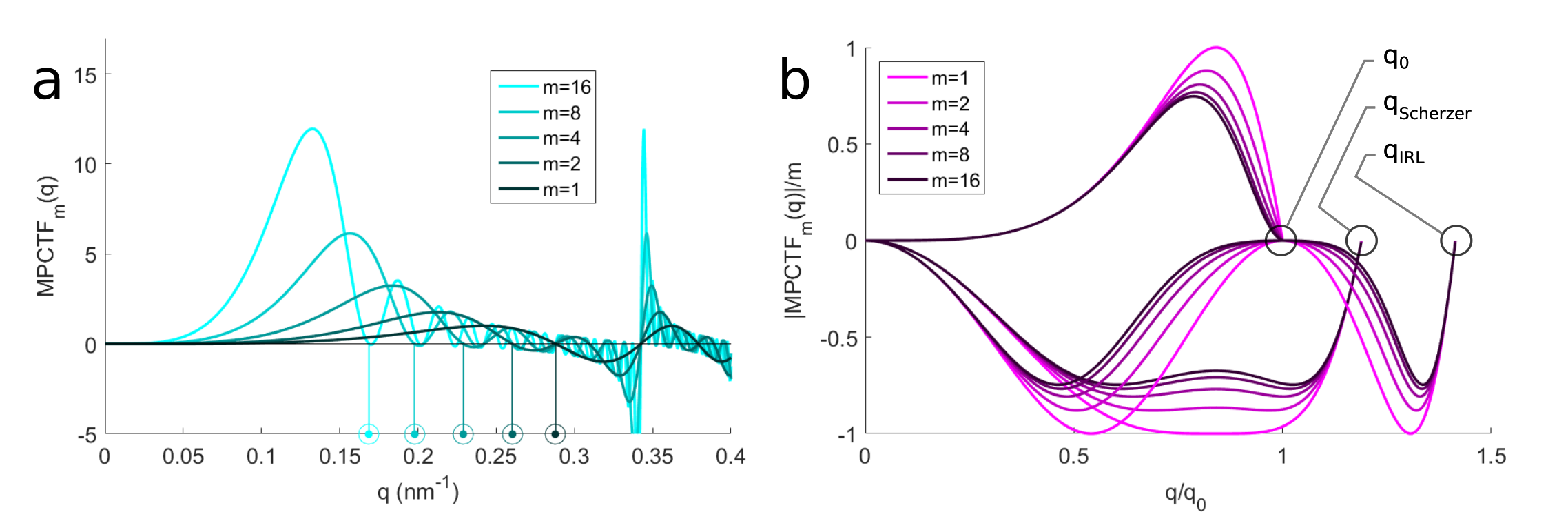}
\caption{a) Multi-pass contrast transfer functions (MPCTFs) with \SI{160}{\milli\metre} spherical aberration for various numbers of passes ($m$). The peak contrast is roughly proportional to $m$ and the point resolution (shown by the lines dropping to the $x$ axis) scales like $m^{-1/4}$. b) MPCTFs normalized by $m$ with 3 different defocus values: no defocus (point resolution $q_0$), $\Delta f_\text{S}$ (point resolution $q_\text{Scherzer}$), and $\Delta f_\text{IRL}$ (point resolution $q_\text{IRL}$), as defined in the text. The horizontal axis is scaled by $q_0$ for each curve.}
\label{fig:mpctf}
\end{figure}

From equation \ref{eq:MPCTF} we see that when $\chi(\vec{q})$ is equal to the $m^{\text{th}}$ root of unity, the contrast transfer function will be zero regardless of the choice of $\phi(\vec{q})$. As a result, adding defocus to the projection optics alone will not be an effective way to increase the point resolution. Point resolution can be increased by adding defocus to the multi-pass optics. In particular, we can define a muli-pass Scherzer defocus \cite{Scherzer1949}, $\Delta f_\text{S}(m)$, as well as the defocus which corresponds to the instrumental resolution limit (IRL, defined as the largest possible first zero of the CTF), $\Delta f_{\text{IRL}}(m)$. These expressions are summarized in table \ref{tab:res}. Compared to zero defocus, $\Delta f_\text{S}(m)$ increases the point resolution by a factor of $\sqrt[4]{2}$ and broadens the high contrast band. Using the multi-pass Scherzer defocus, the point resolution in angular coordinates ($\lambda q$) is \SI{4.2}{\milli\radian} for $m=1$ and \SI{2.7}{\milli\radian} for $m=10$. The designed numerical aperture of \SI{3}{\milli\radian} is a compromise between these values. Adding an additional factor of $\sqrt{2}$ brings the defocus to $\Delta f_{\text{IRL}}(m)$, which increases the resolution by another factor of $\sqrt[4]{2}$ but decreases contrast near the zero-defocus point resolution. Figure \ref{fig:mpctf} shows MPCTFs with various defocus settings.

\begin{table}[]
\begin{tabular}{c|c|c|c|c|c|c|}
\cline{2-7}
 &$\Delta f(m)$ & $\Delta f(1)$ & $\Delta f(10)$ & $p(m)$ & $p(1)$ & $p(10)$ \\\hhline{-::======}
 \multicolumn{1}{|c||}{no $\Delta f$}&0 & 0 & 0 & $\sqrt[4]{\frac{(m+1)\lambda^3C_s}{4}}$ & \SI{3.5}{\nm} & \SI{5.3}{\nm}\\ \hhline{|-||-|-|-|-|-|-|}
 \multicolumn{1}{|c||}{$\Delta f_\text{S}$} &$-\sqrt{\frac{2\lambda C_s}{m+1}}$ & -\SI{1.4}{\micron} & -\SI{600}{\nm} & $\sqrt[4]{\frac{(m+1)\lambda^3C_s}{8}}$ & \SI{2.9}{\nm} & \SI{4.5}{\nm}\\ \hhline{|-||-|-|-|-|-|-|}
 \multicolumn{1}{|c||}{$\Delta f_{\text{IRL}}$}&$-\sqrt{\frac{4\lambda C_s}{m+1}}$ & -\SI{2}{\micron} & -\SI{840}{\nm} & $\sqrt[4]{\frac{(m+1)\lambda^3C_s}{16}}$ & \SI{2.5}{\nm} & \SI{3.8}{\nm}\\ \hhline{|-||-|-|-|-|-|-|}
\end{tabular}
\caption{Expressions and example values (for $m=1$ and 10) for defocus, $\Delta f$, and point resolution, $p$, for three defocus settings: no defocus, $\Delta f_\text{S}$, and $\Delta f_\text{IRL}$, as defined in the text.}
\label{tab:res}
\end{table}

Regardless of the defocus setting, the point resolution scales like $m^{-1/4}$. This is the expected result, since aberrations of sequential elements are additive (if $C_s$ is the spherical aberration of one element, then $mC_s$ is the spherical aberration of $m$ identical elements at constant magnification) and point resolution scales like $C_s^{-1/4}$ \cite{Scherzer1949}. In the future, we expect a high voltage MPTEM will achieve optimal performance with $m\sim10$ \cite{Juffmann2017}. In an aberration-corrected system, this puts the aberration-limited point resolution safely above the dose-limited resolution for proteins \cite{Egerton2014}. Accounting for aberrations from $m\sim$10 passes and also from out-coupling,  the \SI{10}{\kilo\electronvolt} MPTEM will have about 5 nm resolution over a \SI{2}{\micro\metre} field of view.

\section{Transient effects} \label{sec:transient}

In order for the gated mirrors to have stable optical properties, they must behave like electrostatic elements while interacting with the beam. If the \SI{100}{\volt} pulse needed to temporarily switch their behavior is mis-timed or causes ringing in the electrodes, the beam could be mis-focused, deflected, aberrated, or could gain or lose energy from/to the transient fields \cite{Juffmann2015}. Ringing can be prevented by minimizing the high-frequency content of the pulse. The round-trip time in the cavity is between 12 and \SI{20}{\nano\second}, depending on the mirror tuning. The gated mirrors must switch and stabilize in this time, so the fall time (to 1/$e$) of the in-coupling pulse should be less than \SI{4}{\nano\second} and the rise time of the out-coupling pulse should be less than \SI{4}{\nano\second}.

A Gaussian pulse would be simple to generate and would have minimal high frequency content. In order to keep the in-coupling-induced energy spread of the beam below the source spread (0.5 eV), the timing between the source and the voltage pulse must be better than \SI{100}{\pico\second}.

\begin{figure}
\includegraphics[width=1 \linewidth]{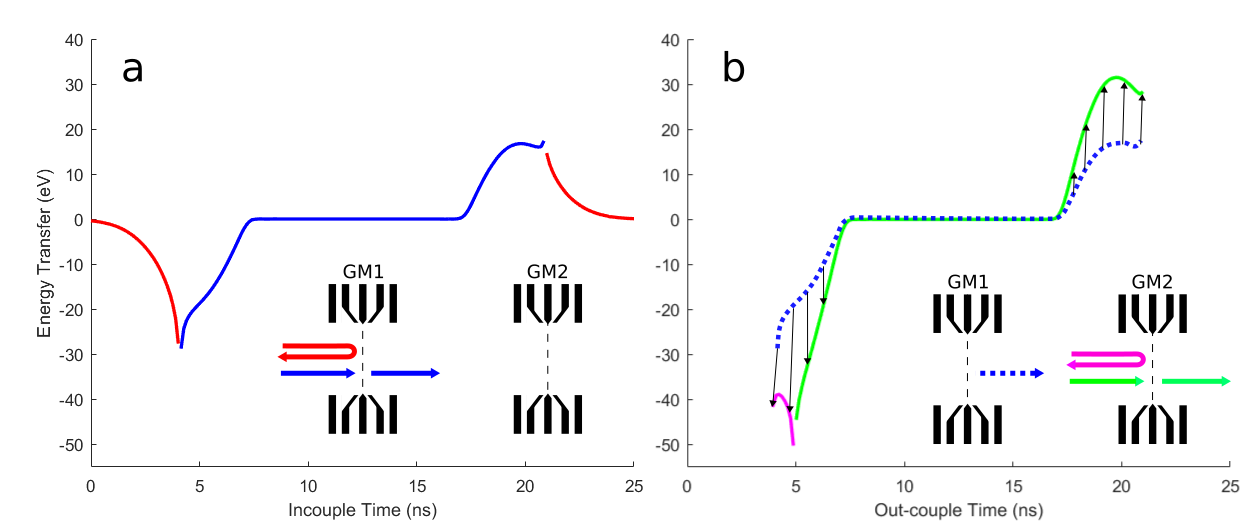}
\caption{Energy transferred to electrons during in- and out-coupling, which depends on their timing relative to a \SI{10}{\nano\second} flat top pulse with \SI{3}{\nano\second} rise time. a) The red curve represents the energy of electrons which are rejected by GM1. The blue curve corresponds to successfully in-coupled electrons. In general, electrons that arrive early lose energy while electron that arrive late gain energy. b) The in-coupled electrons stay in the multi-pass optics for $m=10$ passes, and are then out-coupled. Depending on when they arrive at GM2 relative to the timing of the out-coupling pulse, there may be additional energy transfer. The dotted blue line shows the energies of electrons after in-coupling. The green curve represents electrons which are successfully out-coupled through GM2. The magenta curve represents the electrons which aren't out-coupled.}
\label{fig:incouple}
\end{figure}

A flat-top pulse would relax the timing requirements, but would have more power in high frequency components. Figure \ref{fig:incouple} shows the energy transferred to/from the electron beam depending on the delay between the electron transit and the voltage pulse. Using a continuous source current, about 3/4 of the electrons which are successfully out-coupled have no interaction with transient fields in the mirror. The other 1/4 will experience significantly increased chromatic aberration due to their large energy spread. Using a triggered electron source with less than \SI{10}{\nano\second} jitter removes the additional energy spread from in/out-coupling. 

\begin{figure}
\includegraphics[width=1\linewidth]{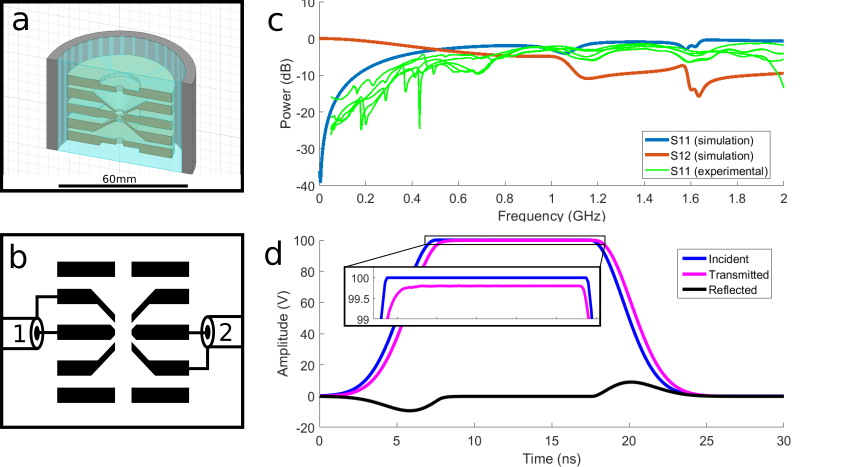}
\caption{a) ANSYS HFSS model of the column near the pulsed mirror electrode. b) Schematic of electrical connections. All electrodes except for the central one are also grounded. c) S-matrix components. The green traces are network analyzer measurements on a prototype mirror using various cable lengths and connections. The consistent features just above \SI{1.1}{\giga\hertz} and \SI{1.6}{\giga\hertz} show qualitative agreement with the simulation. d) Time domain profile of the simulated transmission and reflection of a \SI{10}{\ns} flat top pulse with \SI{3}{\nano\second} rise/fall time (1/$e$). The inset compares the heights and profiles of the incident and transmitted pulse.}
\label{fig:rf}
\end{figure}

We used ANSYS high frequency simulation suite (HFSS) to simulate the transient response of the gated mirror structure to a fast voltage pulse and built a prototype electrode structure to validate the modelling. The results, shown in figure \ref{fig:rf}, show the gated mirror structure has a resonance just above \SI{1}{\giga\hertz}. We acquired reflection data (S11) on the prototype mirror using a vector network analyzer (HP 8510C) with a \SI{50}{\ohm} termination and found qualitative agreement with the simulation. A flat-topped pulse with \SI{3}{\nano\second} Gaussian transitions has negligible spectral content above \SI{1}{\giga\hertz} and thus shows no detectable ringing and negligible reflections in the simulation.

\section{Simulated performance} \label{sec:performance}

\begin{figure}
\includegraphics[width=1 \linewidth]{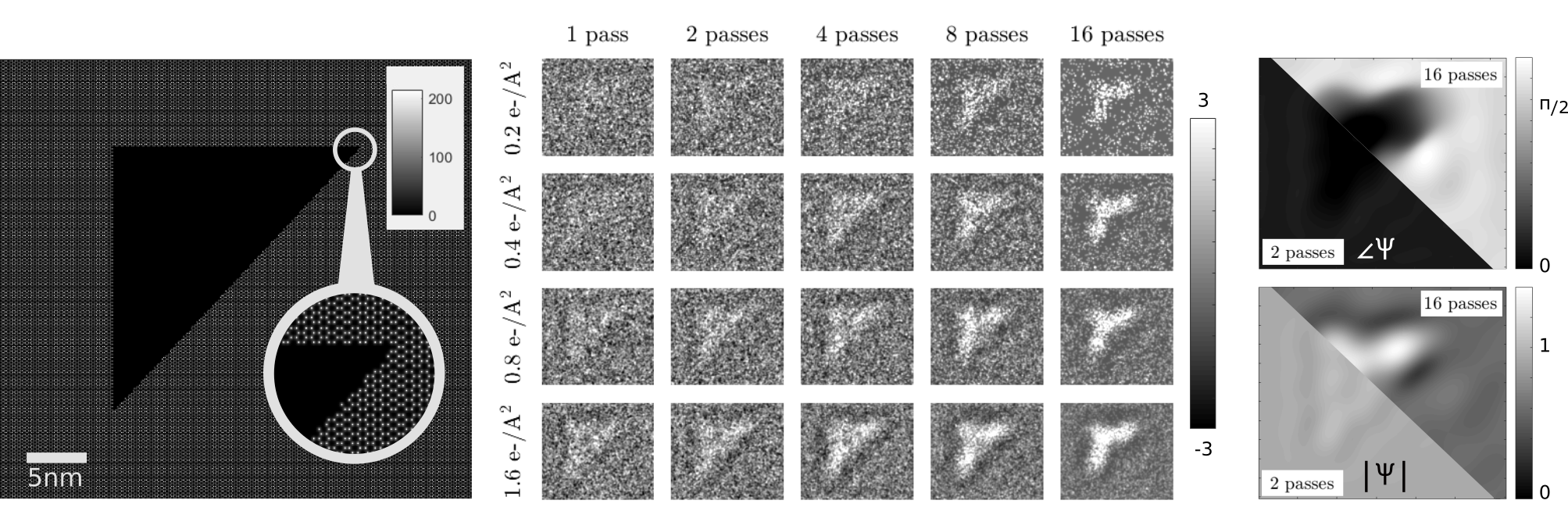}
\caption{Simulation of a triangular hole in single-layer graphene imaged with a \SI{10}{\kilo\electronvolt} MPTEM. Left: sample potential. The color bar shows the projected potential in volt-angstroms. Center: simulated micrographs with various doses and numbers of passes. The number of in-coupled electrons is adjusted along each row so that the total dose (the number of sample-electron interactions) remains constant. The simulation includes 2.8\% inelastic loss and 160 mm spherical aberration per round trip. Phase contrast is generated using \SI{2}{\micron} positive defocus. The color scale is normalized to the standard deviation of the intensity at the detector. Right: exit phase (top) and amplitude (bottom) of the electron wavefunction after 2 and 16 passes.}
\label{fig:graphene}
\end{figure}

An ideal reference sample for proving the MPTEM concept should be as thin as possible to minimize inelastic loss, yet durable enough to survive multiple measurements. Single-layer graphene is a natural choice. 
Figure \ref{fig:graphene} shows simulated images of a \SI{20}{\nano\metre} triangular hole in single-layer graphene (simulation details are described in \cite{Juffmann2017}). The simulation includes \SI{160}{\milli\metre} spherical aberration applied each pass. To estimate the inelastic losses, we use the total inelastic mean free path of graphite at \SI{10}{keV}, $\lambda_{\text{MFP}}=$\SI{12}{\nm} \cite{Shinotsuka2015}, which corresponds to 2.8$\%$ loss per pass (for graphene thickness \SI{0.355}{\nm}). The mean phase shift through the graphene is \SI{0.1}{\radian}. The contrast comes in part from the inelastic losses in the graphene and elastic scattering outside of the \SI{3}{\milli\radian} objective aperture, which is also applied every pass. In addition, \SI{2}{\micron} defocus is applied to generate phase contrast. The defocus must be positive to create positive phase contrast which will add constructively with the amplitude contrast features. Along each row of the micrograph grid, fewer total electrons are used to produce each image as $m$ is increased in order to keep the total dose constant. As $m$ is increased, shot noise decreases more quickly than signal decreases, causing the SNR to rise \cite{Juffmann2017}. We see similar SNR comparing the images along diagonals of the grid. For example, an image formed using 1 pass and dose $d$ has similar contrast to images formed using $m$ passes and dose $d/m$.  

\section{Conclusions}
We have designed a \SI{10}{\kilo\electronvolt} MPTEM column to demonstrate multi-pass electron microscopy. The enabling components are two fast-switching gated mirrors which capture the electron beam in a re-imaging cavity so it can interrogate the sample many times before being transferred to the projection optics. The gated mirror structure was designed to limit electron interactions with the switching fields, but the best performance will be achieved by timing the electrons using a laser-triggered source. The gated mirrors also partially correct aberrations from the objective lenses. The resolution of the microscope is limited by spherical aberration and decreases slowly with the number of passes: the blur from a single pass is about \SI{4}{\nano\metre} and the blur from 10 passes is less than 7 nm. 

\section{Acknowledgements}

This work was done as part of the Quantum Electron Microscope collaboration funded by the Gordon and Betty Moore foundation. We thank Ruud Tromp for discussions about the magnetic prism array and Colin Ophus for the code used to create the graphene imaging simulation. Adam Bowman and Brannon Klopfer acknowledge support from the Stanford Graduate Fellowship.
Adam Bowman acknowledges support from the National Science Foundation Graduate Research Fellowship Program under grant no. 1656518.

\bibliographystyle{unsrt}

\end{document}